\documentclass[preprint,preprintnumbers,amsmath,amssymb]{revtex4}
\usepackage{booktabs}
\usepackage{mathrsfs}
\usepackage{epsfig}
\usepackage{graphicx}
\usepackage{dcolumn}
\usepackage{bm}
\usepackage{amsmath}
\usepackage{slashed}
\usepackage{multirow}

\let\jnfont=\rm
\def\NPB#1,{{\jnfont Nucl.\ Phys.\ B }{\bf #1},}
\def\PLB#1,{{\jnfont Phys.\ Lett.\ B }{\bf #1},}
\def\EPJC#1,{{\jnfont Eur.\ Phys.\ Jour.\ C }{\bf #1},}
\def\PRD#1,{{\jnfont Phys.\ Rev.\ D }{\bf #1},}
\def\PRL#1,{{\jnfont Phys.\ Rev.\ Lett.\ }{\bf #1},}
\def\MPLA#1,{{\jnfont Mod.\ Phys.\ Lett.\ A }{\bf #1},}
\def\JPG#1,{{\jnfont J.\ Phys.\ G}{\bf #1},}
\def\CTP#1,{{\jnfont Commun.\ Theor.\ Phys.\ }{\bf #1},}
\def\ZPC#1,{{\jnfont Z.\ Phys.\ C }{\bf #1},}
\def\JHEP#1,{{\jnfont JHEP \ }{\bf #1},}
\def\Rv{\not{\hbox{\kern-1pt $R$}}}
\def\p{\not{\hbox{\kern-3pt $p$}}}

\newcommand{\beq}{\begin{eqnarray}}
\newcommand{\eeq}{\end{eqnarray}}
\newcommand{\bpmatrix}{\begin{pmatrix}}
\newcommand{\epmatrix}{\end{pmatrix}}

\newcommand{\ba}{\begin{array}}
\newcommand{\ea}{\end{array}}

\newcommand{\be}{\begin{equation}}
\newcommand{\ee}{\end{equation}}

\begin{document}

\title{The Higgs properties in the MSSM after the LHC Run-2 }

\author{ Jun Zhao}
\affiliation{
 Institute of Theoretical Physics, College of Applied Science,
     Beijing University of Technology, Beijing 100124, China;\\
 Institute of Theoretical Physics, Chinese Academy of Sciences, Beijing 100190, China
}

\begin{abstract}
We scrutinize the parameter space of the SM-like Higgs boson in the minimal supersymmetric standard model (MSSM) under current experimental constraints. The constraints are from (i) the precision electroweak data and
various flavor observables; (ii) the direct 22 separate ATLAS searches in Run-1;
(iii) the latest LHC Run-2 Higgs data and tri-lepton search of electroweakinos.
We perform a scan over the parameter space and find
that the Run-2 data can further exclude a part of parameter space.
For the property of the SM-like Higgs boson,
its gauge couplings further approach to the SM values with a deviation below
0.1\%, while its Yukawa couplings $hb\bar{b}$ and $h\tau^+\tau^-$ can still
sizably differ from the SM predictions by several tens percent.

\end{abstract}

\maketitle

\section{Introduction}
\label{sec:intro}
Probing new physics beyond the Standard Model (SM) is the most important task in today's
high energy physics. Among numerous new physics theories, low energy supersymmetry (SUSY)
is the most appealing candidate since it  predicts a light Higgs boson, provides
a candidate for the cosmic dark matter and achieves the unification of gauge couplings.
The SUSY particles (sparticles) have been being intensively searched in the LHC experiments.

Although the discovery of a 125 GeV Higgs boson \cite{higgs-discovery}
 can serve as a good harbinger for SUSY,
the null search results of sparticles at the LHC experiments are continuously squeezing SUSY.
 Especially,  the latest LHC Run-2 searches have further pushed up
the spartcle masses, the current LHC Run-2 measurements for the 125 GeV Higgs boson agree
with the SM predictions and the searches for non-SM Higgs bosons gave negative results.
All these search and measure results should further restrain the parameter space of SUSY.
On the other hand, some precision measurements, such as the precision electroweak data and
various flavor observables, should also play a role in restricting the parameter space
of SUSY.

Using the current experimental data to restrict the SUSY parameter space is a tough
task since we have various SUSY models and some models have too many free parameters
(note that the mass limits on sparticles given by experimental groups are usually
obtained in some over-simplified models).
For the popular minimal supersymmetric standard model (MSSM) \cite{mssm},
we have over one hundred parameters and a comprehensive scan over the parameter space
under current experimental data is rather challenging. Despite of the difficulty, some attempts 
have been tried in the literature \cite{scan1,scan-cmssm,scan2} and some machine learning method
is being explored \cite{new-scan-method}.  
In this work, instead of scanning over the whole parameter space of the MSSM, we focus only
on the Higgs sector, which has a small number of sensitive
parameters (non-sensitive parameters can be fixed).

The structure of this paper is organized as follows.
In Sec.\ref{section2}, we will briefly describe the MSSM by focusing on the Higgs sector.
In Sec.\ref{section3}, we perform a numerical scan over
the parameter space for the Higgs sector.
The properties of the SM-like 125 GeV Higgs boson are demonstrated.
Finally, our conclusion is presented in Sec.\ref{section4}.

\section{The Higgs sector in the MSSM}
\label{section2}
 The MSSM is the most economic realization of SUSY and has a minimal
Higgs sector consisting of two Higgs doublets with opposite hypercharges \cite{mssm}:
\begin{eqnarray}
H_u = \left(\begin{array}{c} H_u^+ \\ H_u^0 \end{array} \right), \qquad
H_d = \left(\begin{array}{c} H_d^0 \\ H_d^- \end{array} \right) \, .
\end{eqnarray}
The tree-level Higgs potential is given by
\begin{eqnarray}
V & = & m_1^2 |H_u|^2 + m_2^2 |H_d|^2 - B_\mu \epsilon_{\alpha\beta} (H_u^\alpha H_d^\beta
+ h.c.) \nonumber \\
& & +\frac{g^2+g'^2}{8} (|H_u|^2-|H_d|^2)^2 + \frac{g^2}{2}
|H_u^\dagger H_d|^2 \; ,
\end{eqnarray}
with $\epsilon_{\alpha\beta}$ being the antisymmetric tensor, $g$ and $g'$ are the
SM SU(2) and U(1) gauge couplings.
A typical feature of this potential is that the quartic Higgs couplings are fixed
by gauge couplings and hence a light rather Higgs boson is predicted.
With spontaneous breaking of electroweak symmetry, the neutral components
of the two Higgs doublets both develop vacuum expectation values $v_{u,d}$, whose
squared sum is $v_u^2+v_d^2=v^2$ with $v\approx 246$ GeV and their ratio
$\tan\beta=v_u/v_d$ is a free parameter.
A neutral Goldstone $G^0$ from the neutral components and a pair of charged
Goldstones $G^\pm$ from the charged components are eaten by gauge bosons
$Z$ and $W^\pm$, respectively. The remained degrees of freedom give five
mass eigenstates: $h$, $H$, $A$ and $H^\pm$, with  $h$ and $H$ ($m_h<m_H$) being
CP-even while  $A$ being CP-odd. The lightest one, $h$, is the so-called SM-like
125 GeV Higgs, whose property has been being measured at the LHC experiments
and will be precisely examined at the future CEPC or FCC-ee collider.
Also, the non-SM Higgs bosons, $H$, $A$ and $H^\pm$, have been
being searched at the LHC and their masses are continuously pushed up by the
null search results. In our analysis in the proceeding section
we will use all the relevant Higgs data to constrain the parameter space of
the Higgs sector.

Note that for the MSSM Higgs sector the tree-level potential is not sufficient and
the loop contributions must be considered. Actually, sizable loop effects from
the stops are necessary to enhance the SM-like Higgs mass to 125 GeV.
To achieve this, rather heavy stops or a large trilinear coupling $A_t$ is needed,
which cause the so-called little hierarchy problem for the MSSM. Some extensions of
the MSSM, e.g., the next-to-minimal supersymmetric standard model (NMSSM),
can solve this problem. So in light of the 125 GeV Higgs discovery, the MSSM is no longer
perfectly natural while the NMSSM seems to be more favored  \cite{scan1}.
In our analysis for the MSSM Higgs sector, we will consider the latest results of
loop effects.

We note that currently the MSSM is not so perfect. Besides the little hierarchy problem
caused by the 125 GeV Higgs boson mass, it suffers from an old problem called $\mu$-problem
since a low value of $\mu$ cannot be explained in the framework of MSSM. In addition, the MSSM
has over 100 free parameters because no boundary conditions for the soft parameters are assumed
in this model. So the MSSM is kind of low-energy effective or phenomenological model.
The fancier models like the mSUGRA (minimal supergravity) or CMSSM (constrained MSSM)
are more predictive because they assume boundary conditions for the soft parameters and
have much smaller numbers of free parameters. These models, however, face the problem
of consistency between 125 GeV Higgs boson mass and muon $g-2$ explanation \cite{cmssm-after-125},
which might be tackled with some extensions like the generalized gravity mediation or
deflected anomaly mediation \cite{wangfei}.  
Since the MSSM has so many free parameters, it is almost impossible to scan over
its whole parameter space. Fortunately, for the  Higgs sector,
the sensitive parameters are rather limited in number. In our following analysis we
will perform a scan over the relevant parameter space under current experimental data.

\section{SM-like Higgs under current data }
\label{section3}
\subsection{A scan over the  parameter space}
In our calculations the SM parameters like the masses of gauge bosons, top and bottom quarks
are taken from the Particle Data Book \cite{pdg}.
The parton distribution functions are from \textsf(NNPDF30\_lo\_as\_0130) \cite{pdf}.

In our scan over the relevant parameter space,
we assume all soft parameters are real and the first two generations  of  squarks
or sleptons are degenerate in masses.
The scanned parameters are in the following ranges
\begin{eqnarray}
	&&
	90{\rm ~GeV}\leq
	(m_{\tilde{L}_1} = m_{\tilde{L}_2}, m_{\tilde{e}_1} = m_{\tilde{e}_2},  m_{\tilde{L}_3},  m_{\tilde{e}_3} )
	\leq4 {\rm ~TeV},  \nonumber\\
	&&
	200{\rm ~GeV}\leq
	(m_{\tilde{Q}_1} = m_{\tilde{Q}_2},  m_{\tilde{u}_1} = m_{\tilde{u}_2}, m_{\tilde{d}_1} = m_{\tilde{d}_2})
	\leq4 {\rm ~TeV},  \nonumber\\
	&&
	100{\rm ~GeV}\leq
	(m_{\tilde{Q}_3}, m_{\tilde{u}_3}, m_{\tilde{d}_3})
	\leq4{\rm ~TeV}, \nonumber \\
	&&
	0{\rm ~GeV}\leq |M_1| \leq 4{\rm ~TeV}, \quad 70{\rm ~GeV}\leq |M_2| \leq 4{\rm ~TeV},
	\nonumber\\
	&& 80{\rm ~GeV}\leq |\mu| \leq 4{\rm ~TeV},
	\quad 200{\rm ~GeV}\leq M_3 \leq 4{\rm ~TeV},\nonumber\\
	&&
	0{\rm ~GeV}\leq |A_t| \leq 8{\rm ~TeV},
	\quad 0{\rm ~GeV}\leq (|A_b|, |A_{\tau}|) \leq 4{\rm ~TeV}, \nonumber \\
	&&
	100{\rm ~GeV}\leq M_A \leq 4{\rm ~TeV},\quad  1\leq \tan\beta \leq 60,
\end{eqnarray}
where $\tilde{L}$ ($\tilde{Q}$) denote a left-handed slepton (squark),
$\tilde{e}$ ($\tilde{u}$ or $\tilde{d}$) denote a right-handed slepton (squark),
and $A_t$,  $A_b$ and $A_\tau$ are the tri-linear couplings of stops, sbottoms and staus, respectively.

In our scan we take into account the following constraints at 95\% CL or $2\sigma$ level:
\begin{itemize}
	\item[(1)]Various indirect constraints from the precision electroweak data \cite{Altarelli}, the flavour observables  $b\to s\gamma$ ($(2.99\sim 3.87)\times 10^{-4} $) and $B_s\to \mu^+\mu^-$  ($(1.5\sim 4.3)\times 10^{-9} $) \cite{b-decay}, as well as the muon $g-2$  ($(10.1\sim 42.1)\times 10^{-10} $)\cite{g-2}.
	
	\item[(2)] The LHC Run-1 and Run-2 data on 125 GeV Higgs boson \cite{run1-125-data,run2-125-data}
	and the current search results of non-SM Higgs bosons ($H,A,H^\pm$)\cite{run2-H-A-data}.
	The Higgs search data from LEP and Tevatron are also considered \cite{higgsbounds}.
	We use SusHi-1.6.1 \cite{sushi} to calculate the production rate of non-SM Higgs bosons.
	
	\item[(3)] The LHC Run-1 data from the direct 22 separate ATLAS searches \cite{atlas-22}, including, e.g.,
	$0\text{-}lepton + 2 \text{-}10 jets + E_T^{miss} $,  $1\text{-}lepton +  jets + E_T^{miss}$, monojet
	and stop in various channels.
	
	\item[(4)] The LHC Run-2 data from the CMS search of
	$3\ell(e\ \mathrm{or}\ \mu)+0\mathrm{jet}+\slashed E_T$ \cite{CMS3lepton},
	in which two leptons form an OSSF (opposite sign same flavor) pair. Such a signal can come from
	the production of $\tilde\chi_1^\pm \tilde\chi_2^0$ followed by the decays
	$\tilde\chi_1^\pm \to \tilde\chi_1^0 W)$ and $\tilde\chi_2^0 \to \tilde\chi_1^0 Z)$.
	We use the package \textsf{CheckMate 1.2.2}\cite{CheckMate} to recast the
	LHC $3\ell(e\ \mathrm{or}\ \mu)+0\mathrm{jet}+\slashed E_T$ inclusive process.

	\item[(5)] The dark matter constraints from LUX-2016 data on the spin-independent cross section
	\cite{lux-2016} and relic density from Planck measurement ($0.1146\sim 0.1226$) \cite{planck}.
	We use \textsf{MicrOmega}\cite{micromega} in our calculations.
\end{itemize}
These constraints are listed in Table \ref{constraints}.

\begin{table}[ht!]
	\caption{A list of constraints considered in our scan.}
	\begin{tabular}{|c|l|}
		\hline
		indirect    &~~ precision EW data, invisible $Z$-decay, muon $g-2$ \\
		constraints &~~ falvor observables ( $b\to s\gamma$, $B_s\to \mu^+\mu^-$, $\cdots$)\\ \hline
		LEP \& Tevatron   &~~ non-SM Higgs bosons $H$, $A$, $H^\pm$ \\
		searches           &~~  sparticles (charginos, neutralinos, sleptons) \\ \hline
		LHC Run-1 22 searches &~~ data from ATLAS 22 separate searches (see \cite{atlas-22}) \\ \hline
		LHC Run-1 Higgs data &~~ totally 89 observables (see \cite{run1-125-data}) \\ \hline
		LHC Run-2 Higgs data &~~ totally 30 observables (see \cite{run2-125-data})\\ \hline
		LHC Run-2 non-SM Higgs searches &~~ CMS search of non-SM Higgs bosons $H$ and $A$ \\\hline
		LHC Run-2 tri-lepton data &~~ CMS search of $3\ell(e\ \mathrm{or}\ \mu)+0\mathrm{jet}+\slashed E_T$ \\\hline
		LUX-2016 \& Planck data  &~~ limits on dark matter scattering cross section \\
		on dark matter &~~ dark matter relic density \\
		\hline
	\end{tabular}
	\label{constraints}
\end{table}

\subsection{SM-like Higgs boson under current experimental data}
\begin{figure}[ht!]
  \centering
  \includegraphics[width=6cm]{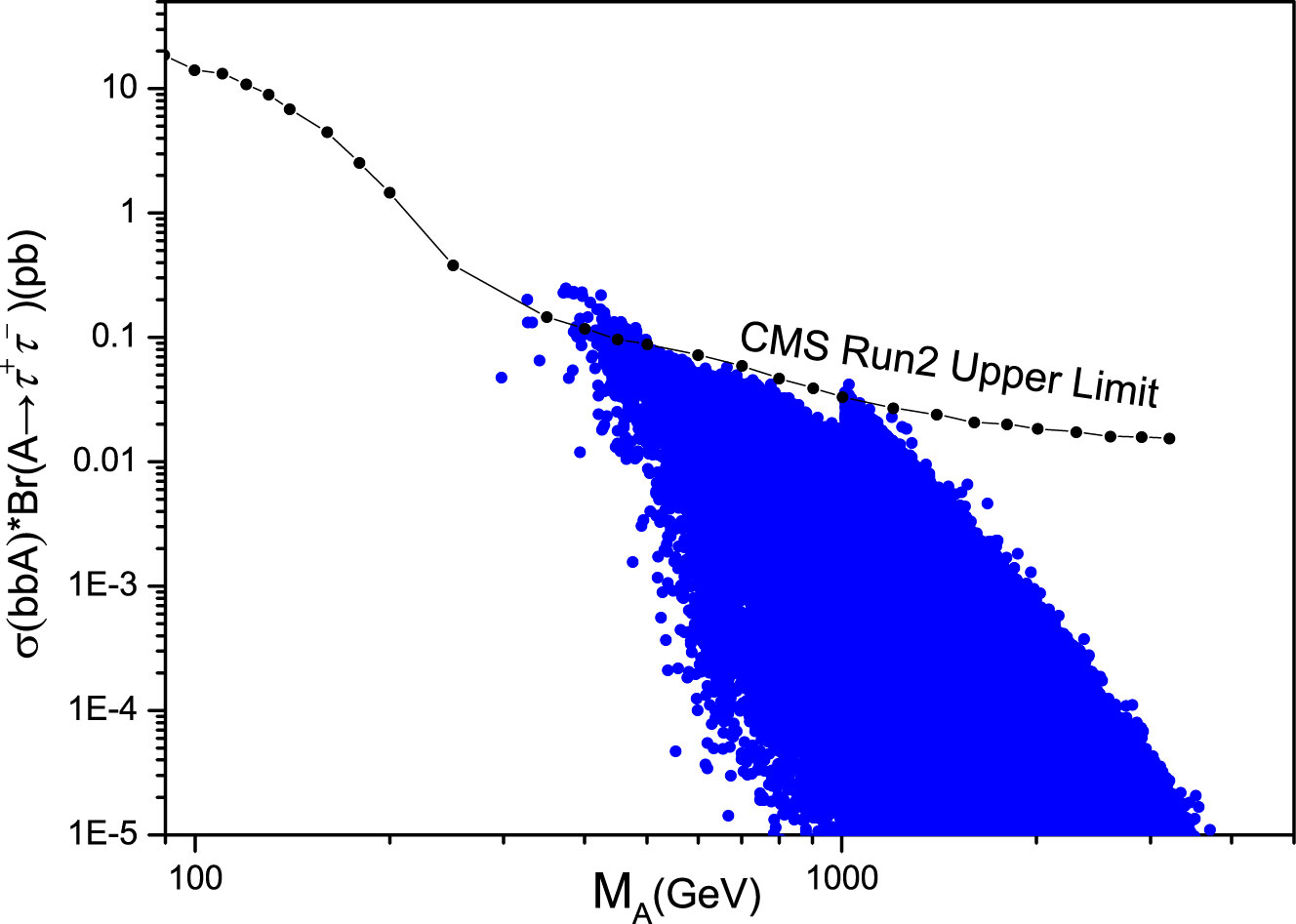}
  \includegraphics[width=6cm]{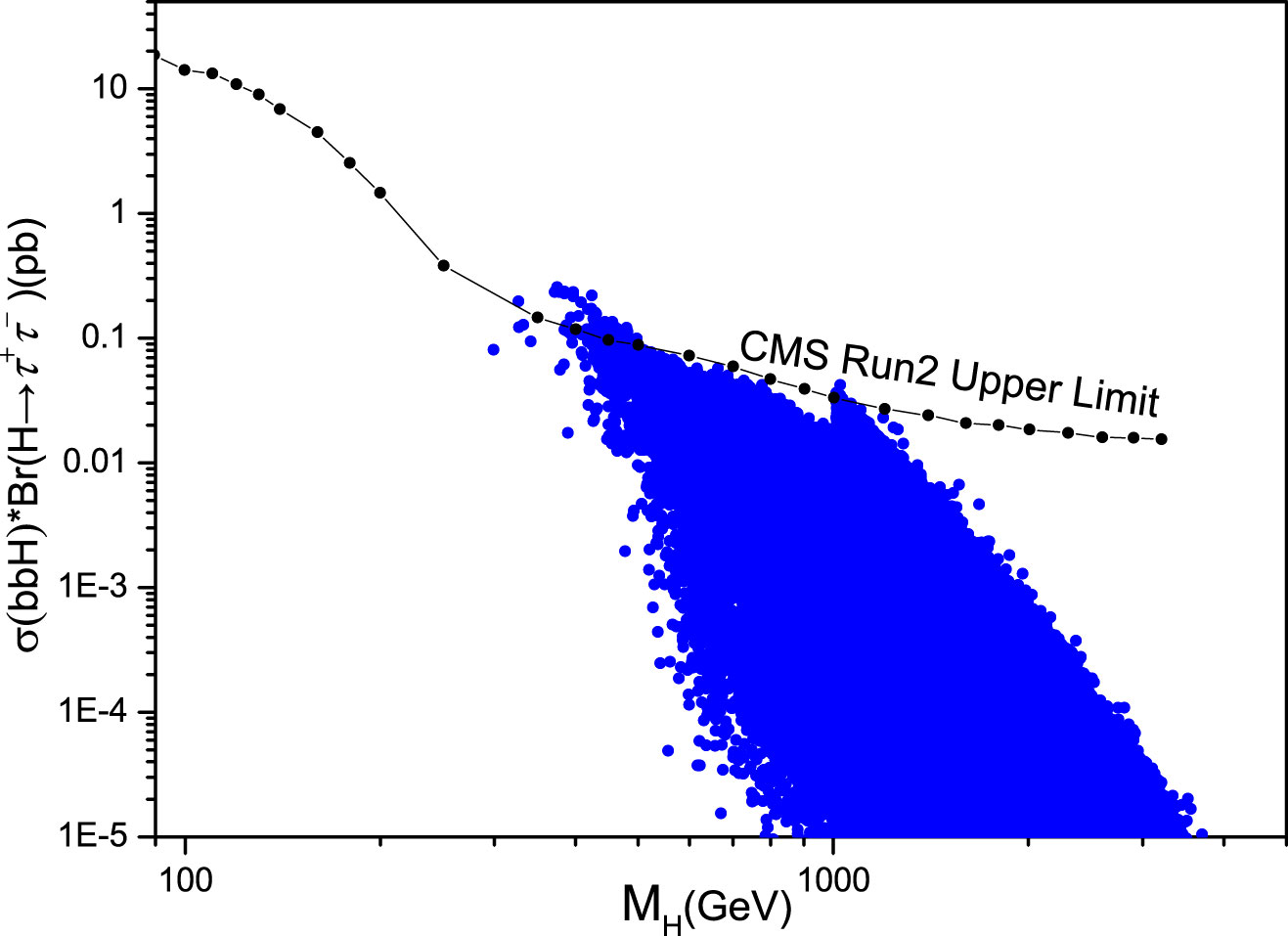}\\
  \includegraphics[width=6cm]{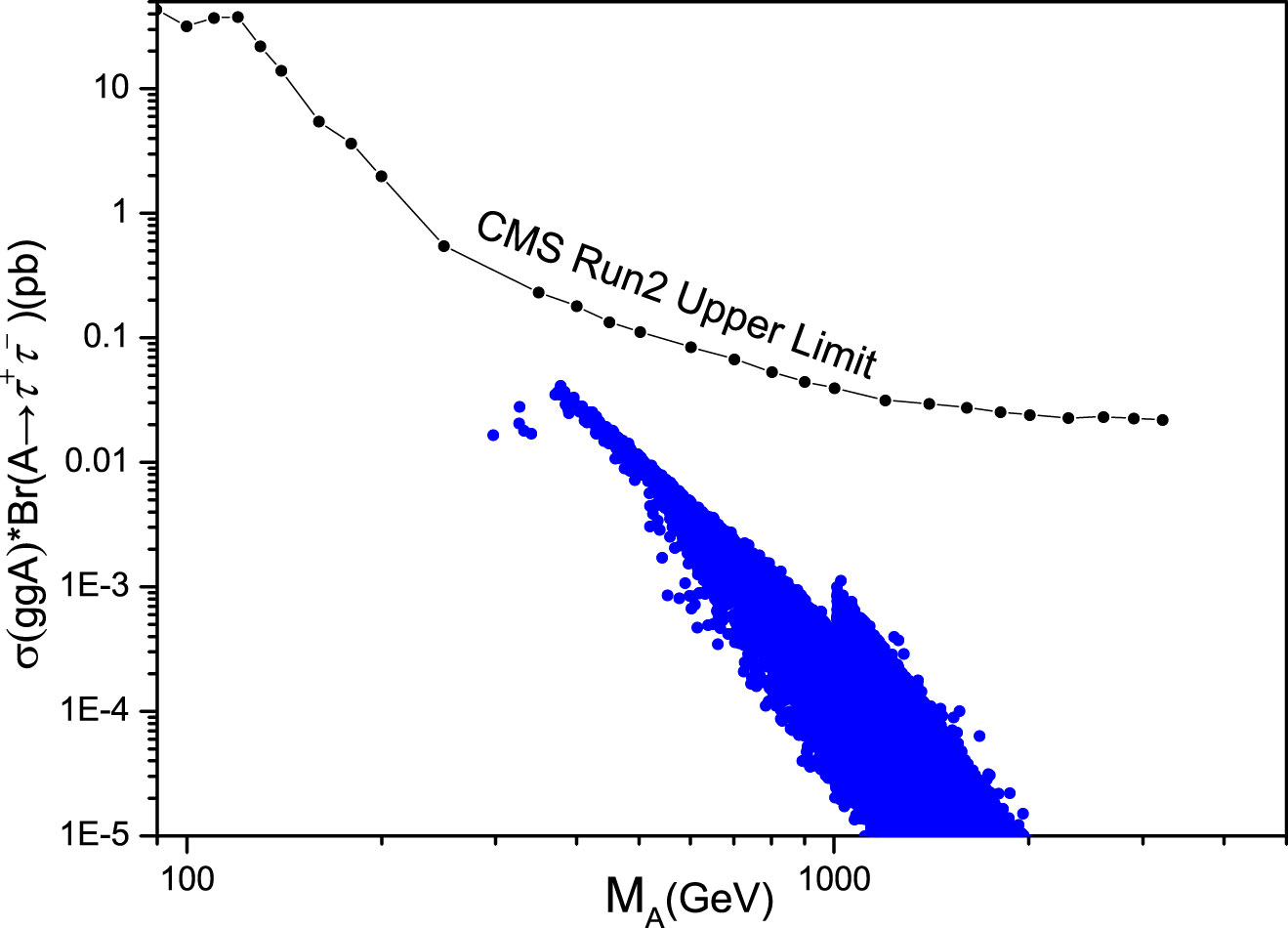}
  \includegraphics[width=6cm]{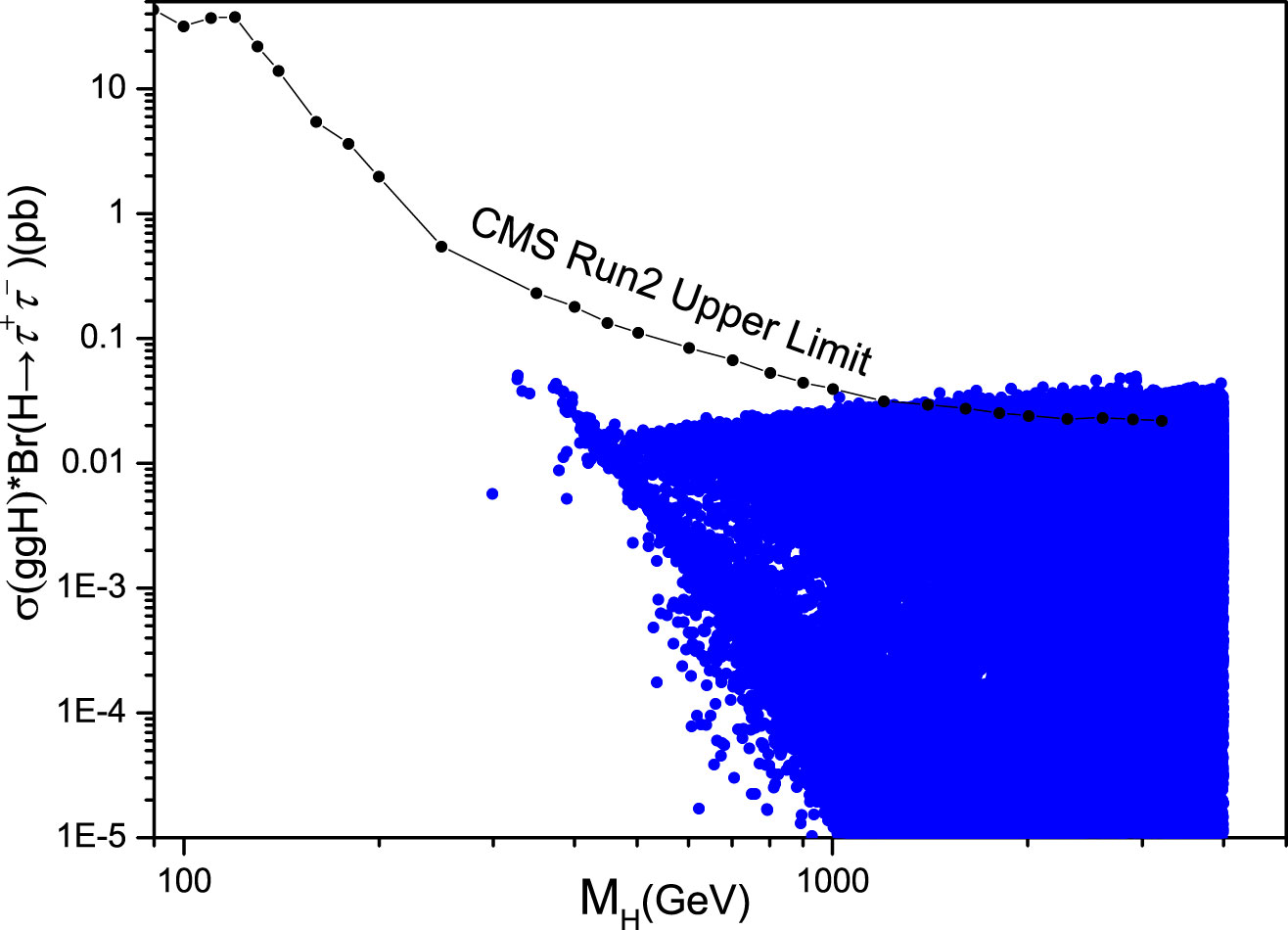}
\vspace{-.5cm}
\caption{Scatter plots of survived samples. The curves are the
upper limits from the current Run-2 data of CMS searches of non-SM Higgs bosons \cite{data-CMSrun2} .}
\label{higgs-fig1}
\end{figure}

In Fig.\ref{higgs-fig1} we display the survived samples to show the current
Run-2 constraints from the CMS searches of non-SM Higgs bosons, which include
the productions of $bbA$, $bbH$, $ggA$ and $ggH$
followed by $A \to \tau^+\tau^-$ and $H \to \tau^+ \tau^-$.
We see that so far only a mall portion of parameter space has been
excluded by such non-SM Higgs searches, owing to the limited luminosity
and tau identification efficiency. Of course, at the end of Run-2, the
sensitivity will be much better due to a higher luminosity.

%Higgs fig.1 %%%%%%%%%%%%%%%%%%%%%%%%%

Next we use the survived samples to fit the LHC Run-1 and Run-2 Higgs data.
In Table \ref{chi2} we present the best sample with minimal $\chi^2$ value
min($\chi^{2}$)/DOF where $\chi^2$ for an observable is defined as the difference of
its theoretical value and experimental value divided by the $1\sigma$ error while
DOF denotes the degree of freedom defined as the number of observables minus the
number of free parameters.
So we see so far the MSSM best point can fit the LHC Higgs data quite well.

%%%%%table   begin%%%%%%%%%%%%%%%
\begin{table}
\caption{The minimal $\chi^2$ values min($\chi^{2}$)/DOF of LHC 125 GeV Higgs data.}
   \setlength{\tabcolsep}{2pt}
  \centering
  \begin{tabular}{|c|c|c|c|}
    \hline
     $    $&~~$Run-1$~~& ~~$Run-2$~~ &~~$Run-1+Run-2$~~\\
    \hline
     ~~${\rm min(\chi^{2})/DOF}$~~
     & $74.4/70  $ & $ 40.3/11$
     & $116.3/100$ \\
     \hline
     \end{tabular}
\label{chi2}
\end{table}

\begin{figure}[ht!]
  \centering
  \includegraphics[width=5.4cm]{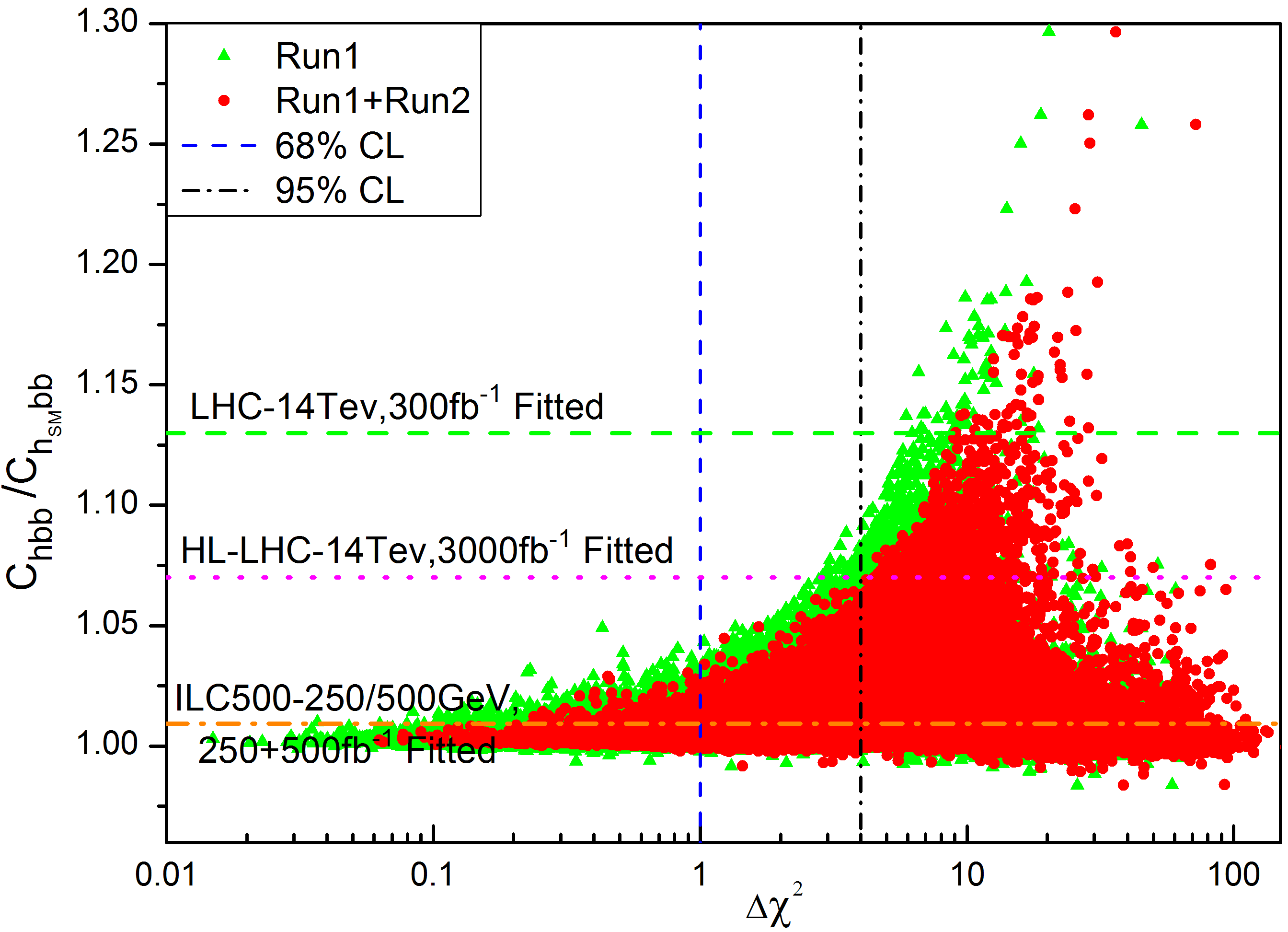}
  \includegraphics[width=5.4cm]{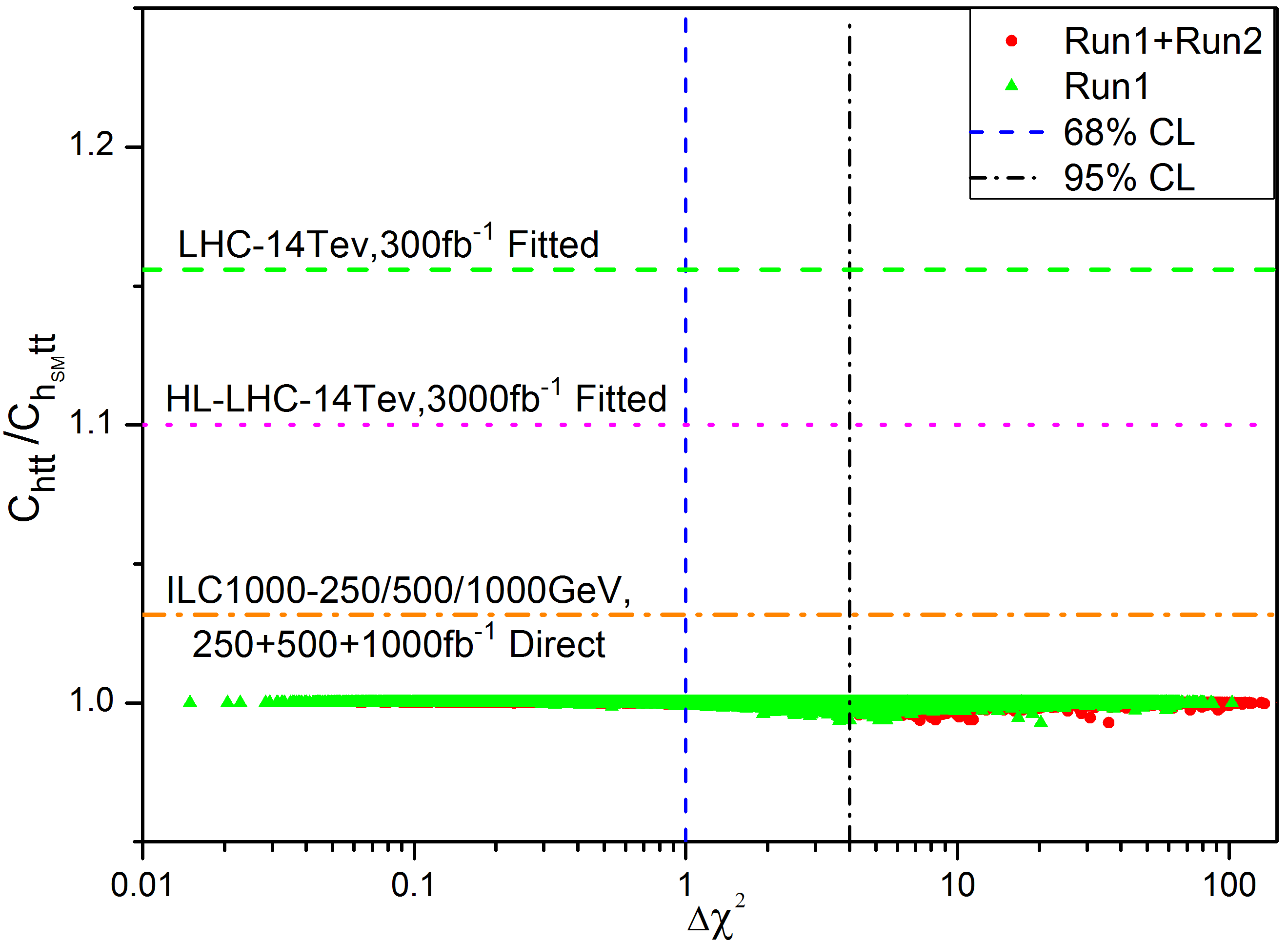}
  \includegraphics[width=5.4cm]{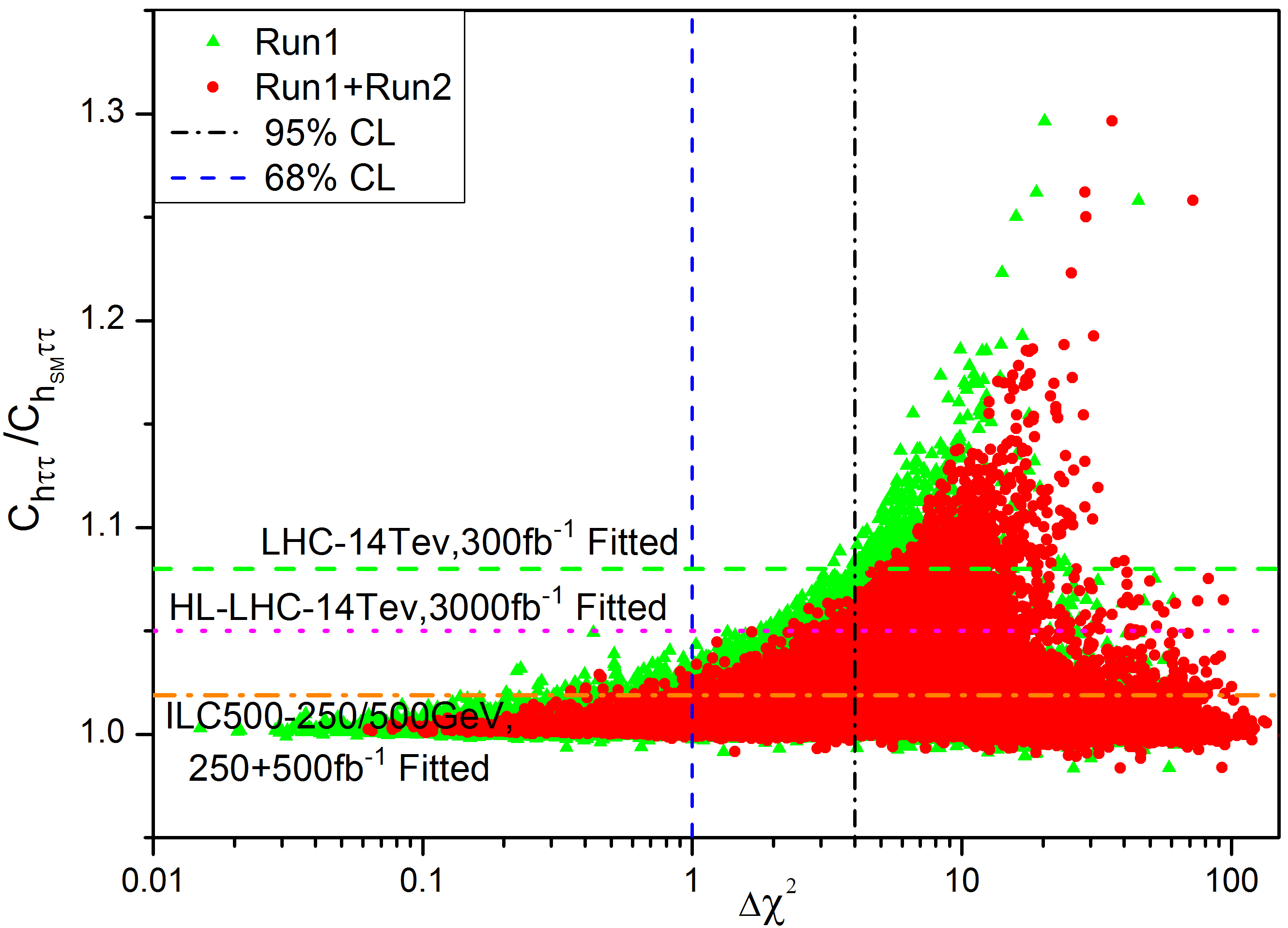}\\
  \vspace{3mm}
 \includegraphics[width=5.4cm]{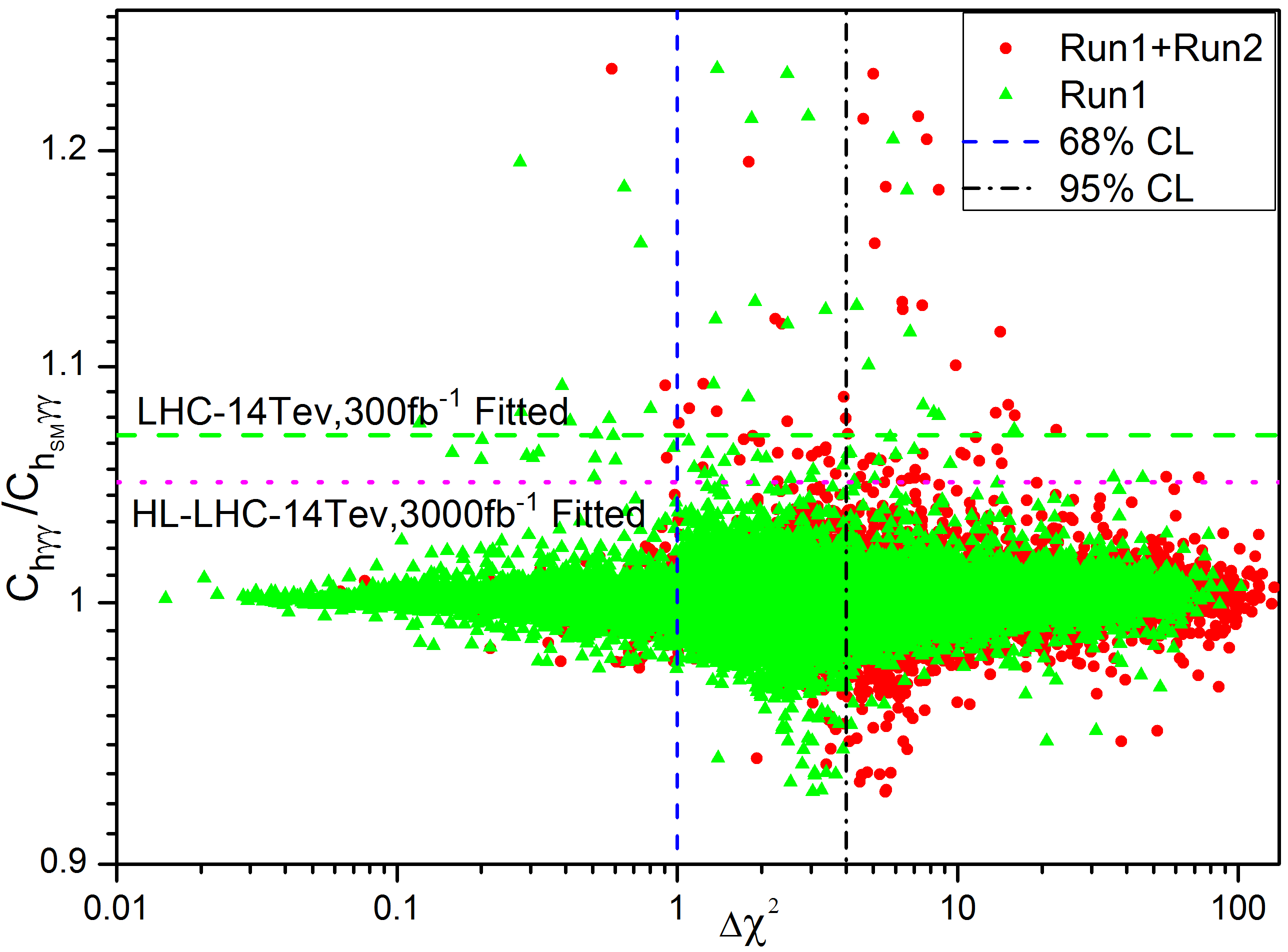}
  \includegraphics[width=5.4cm]{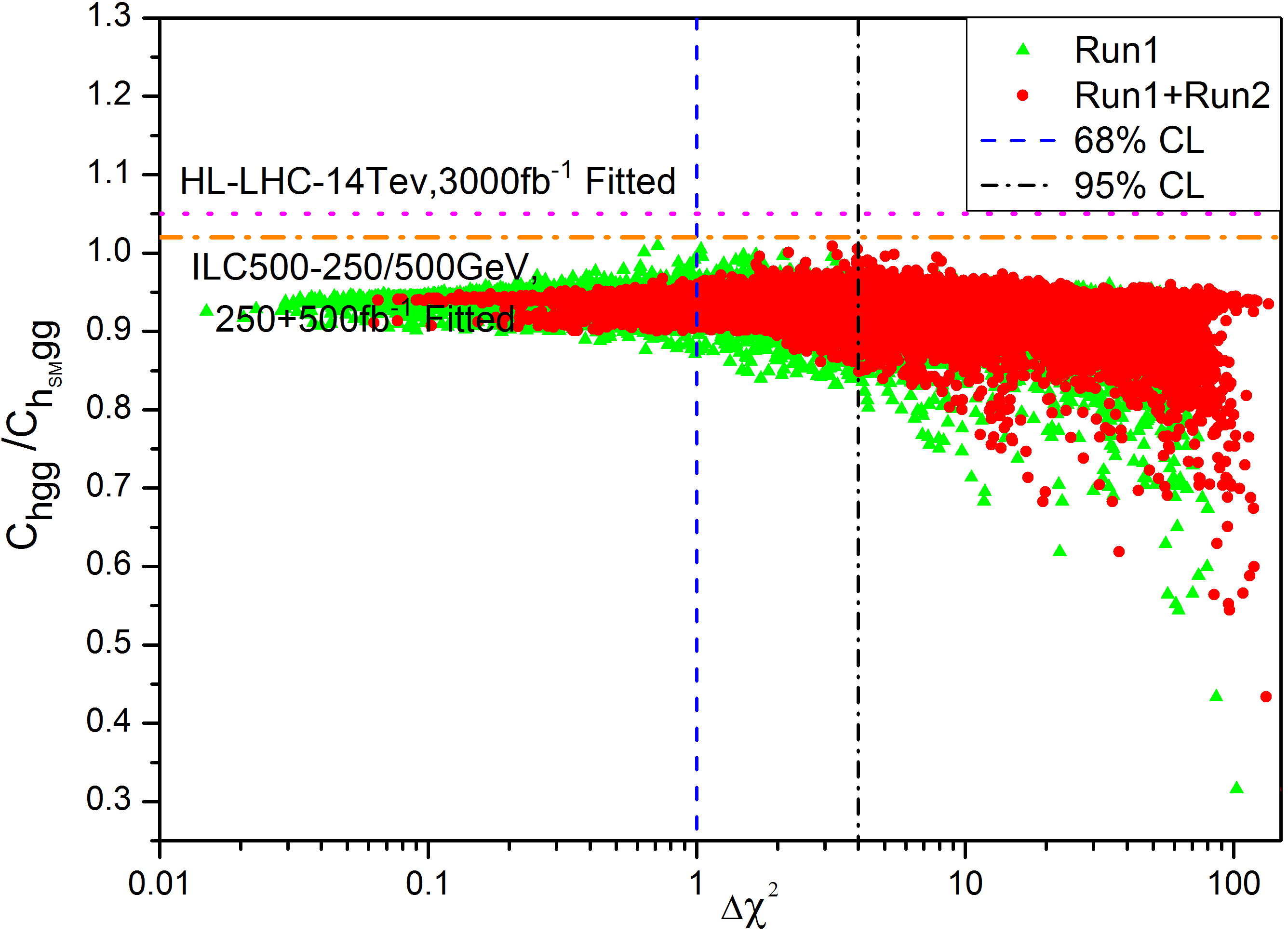}
  \includegraphics[width=5.4cm]{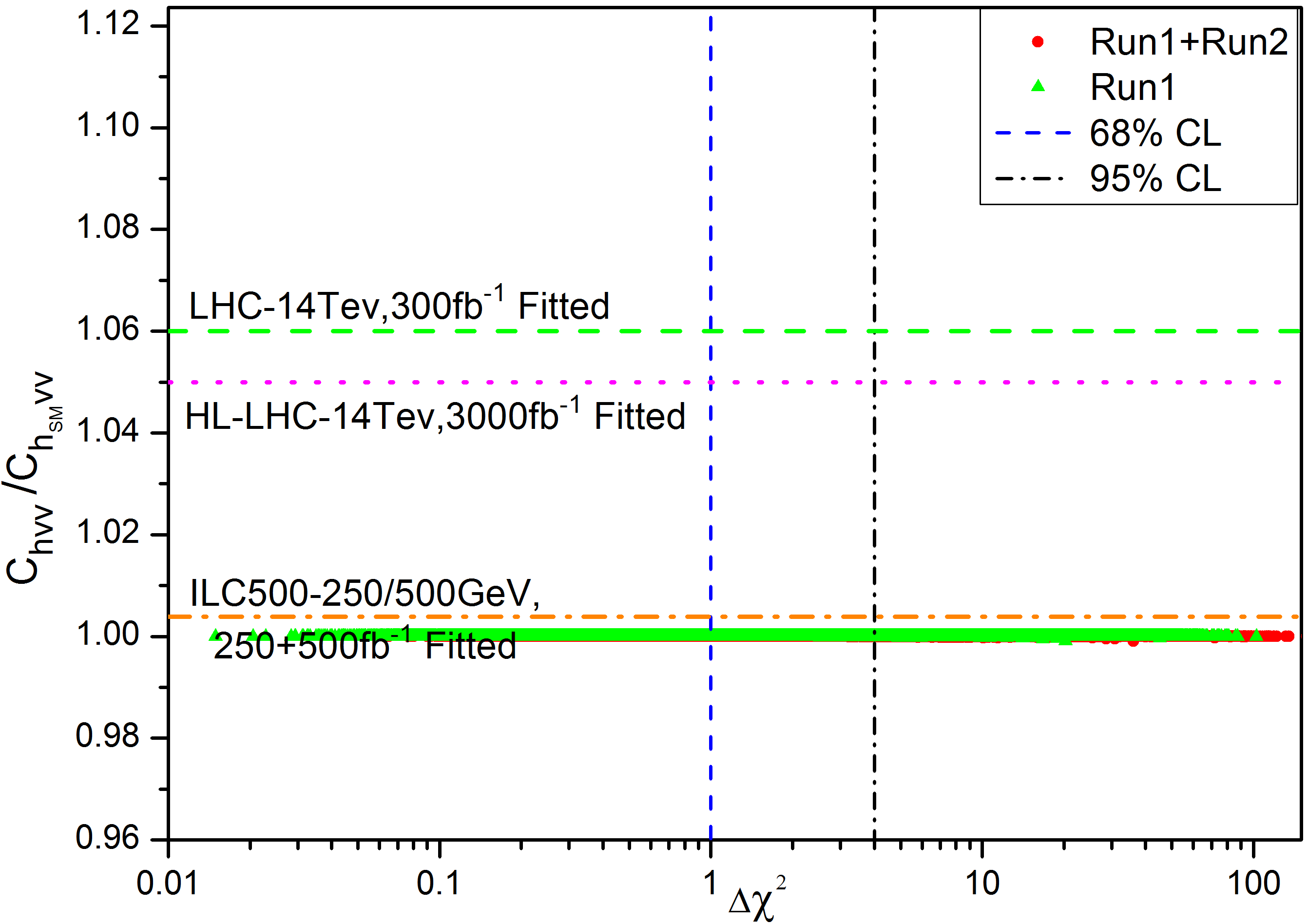}
\vspace{-1.0cm}
\caption{Scatter plots of survived samples showing the $\Delta\chi^2$ values 
(with Run-1 data or Run-1+Run-2 data) for the SM-like Higgs couplings normalized to the SM values.
 The sensitivities of the LHC, HL-LHC and ILC to the Higgs couplings\cite{LHC-ILC}
are also shown.}
\label{higgs-fig2}
\end{figure}

Then in Fig.\ref{higgs-fig2} we show 
$\Delta\chi^2=\chi^2-{\rm min}(\chi^2)$, with Run-1 data or Run-1+Run-2 data, 
for the SM-like Higgs couplings (normalized to the SM values) in the allowed parameter space.
Here we see that gauge couplings of the SM-like Higgs boson
approach to the SM values with a deviation below
0.1\%, while its Yukawa couplings $hb\bar{b}$ and $h\tau^+\tau^-$ can still
sizably differ from the SM predictions by several tens percent.
The deviation of $ht\bar{t}$ coupling is below 2\%.
The loop-induced $hgg$ and $h\gamma\gamma$ couplings can differ from the SM values
significantly by several tens percent.
Note that the $ht\bar{t}$ and  $hgg$ couplings are dropped relative to the SM values,
 the $hb\bar{b}$ and $h\tau^+\tau^-$ couplings are enhanced, while the $h\gamma\gamma$
coupling can be either enhanced or dropped.
The sensitivities of the HL-LHC and ILC to the Higgs couplings are also shown.
We see that a large portion of the parameter space can be covered by ILC or
HL-LHC through measuring the $hb\bar{b}$  and $h\tau^+\tau^-$ couplings \cite{suwei}.

\begin{figure}[ht!]
  \centering
  \includegraphics[width=5.4cm]{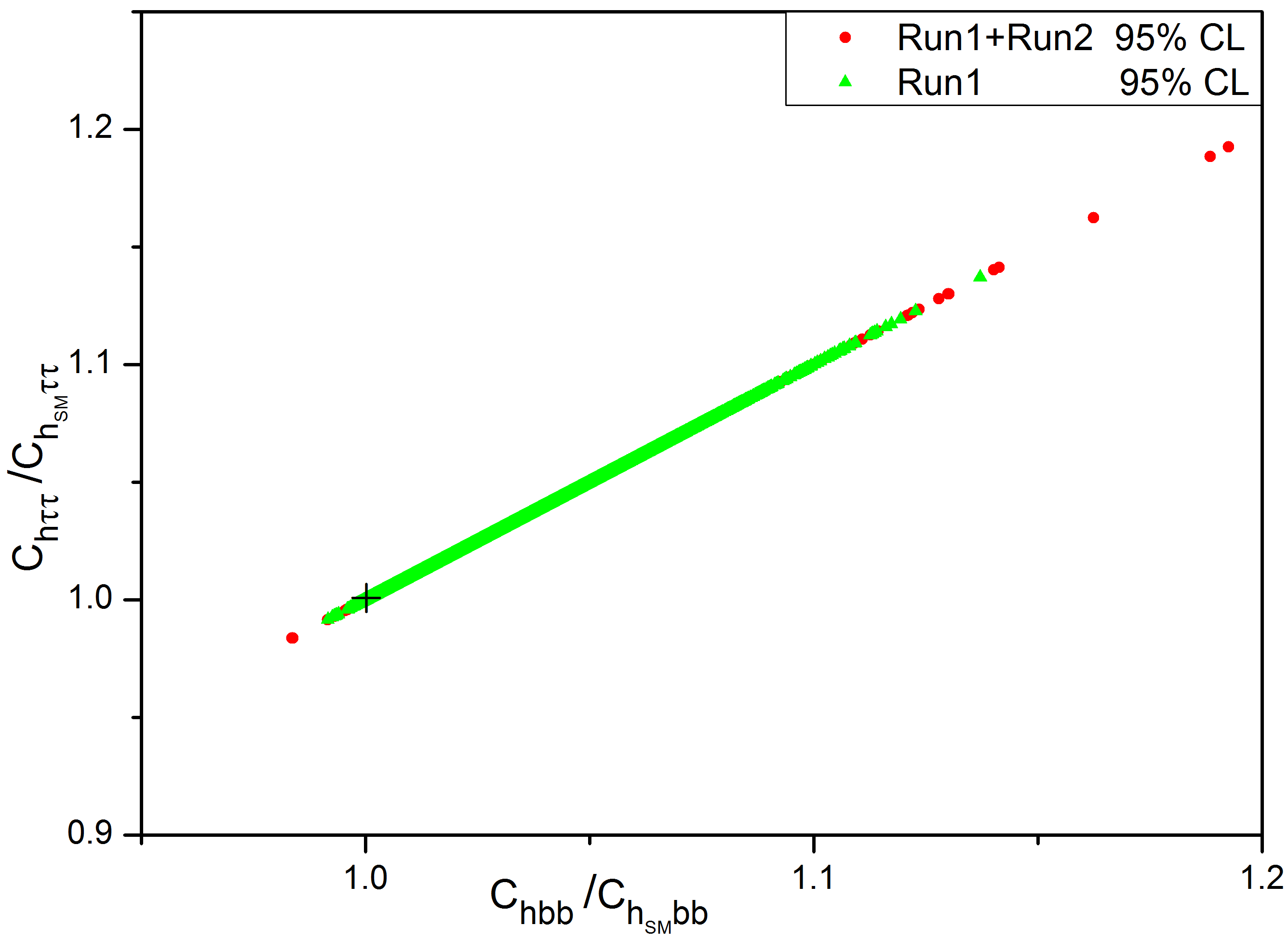}
  \includegraphics[width=5.4cm]{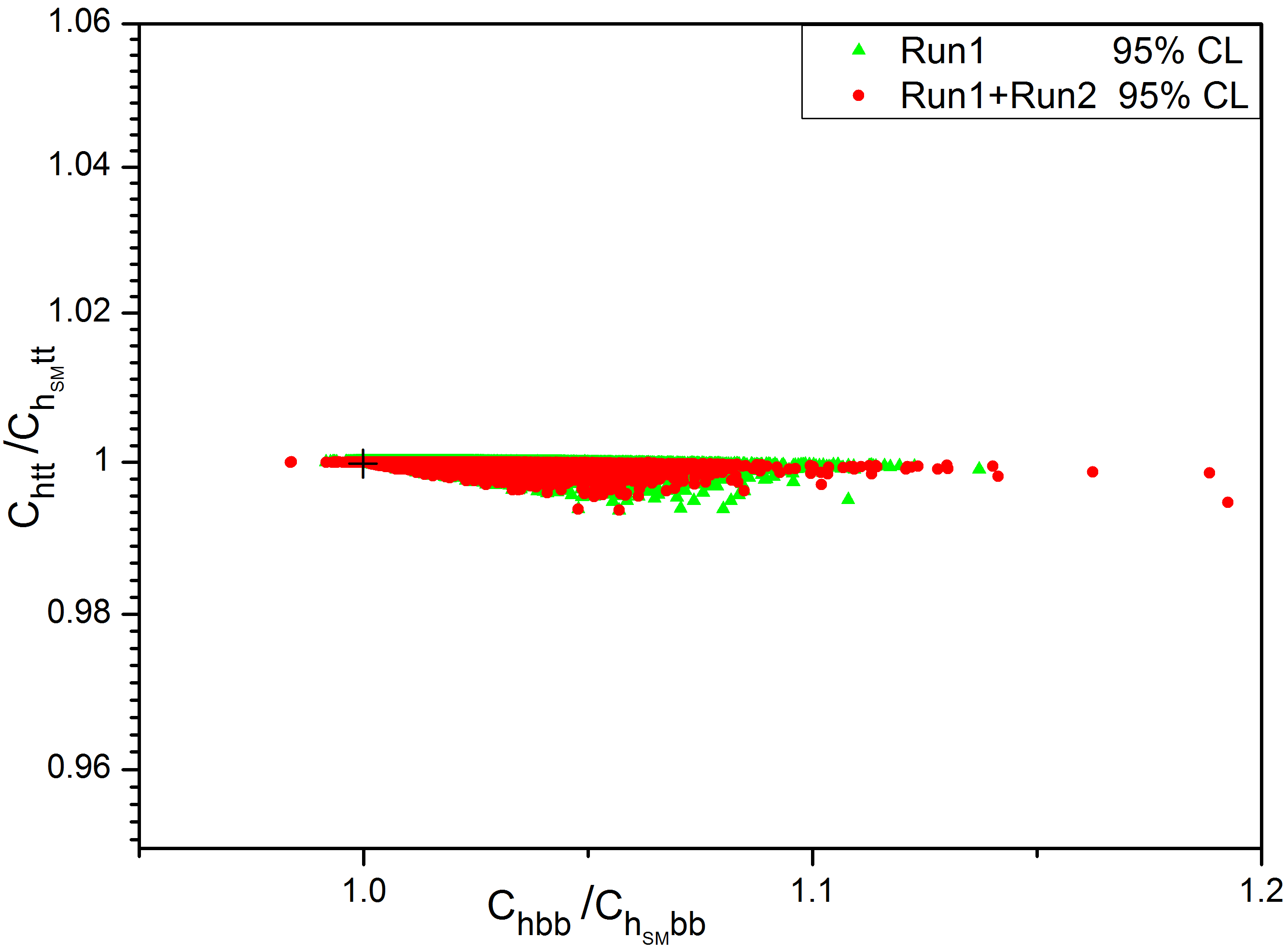}
  \includegraphics[width=5.4cm]{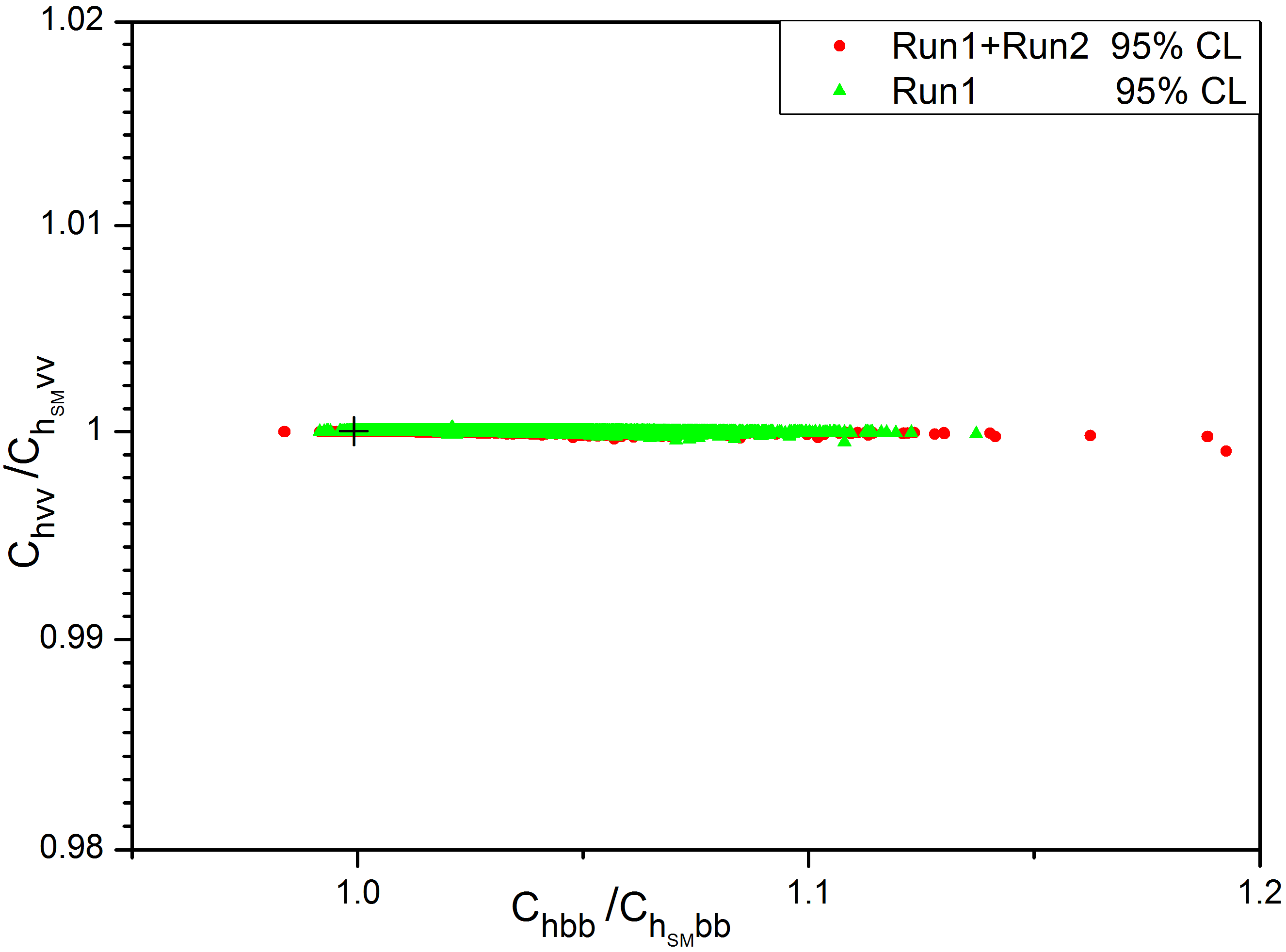}\\
  \vspace{3mm}
  \includegraphics[width=5.4cm]{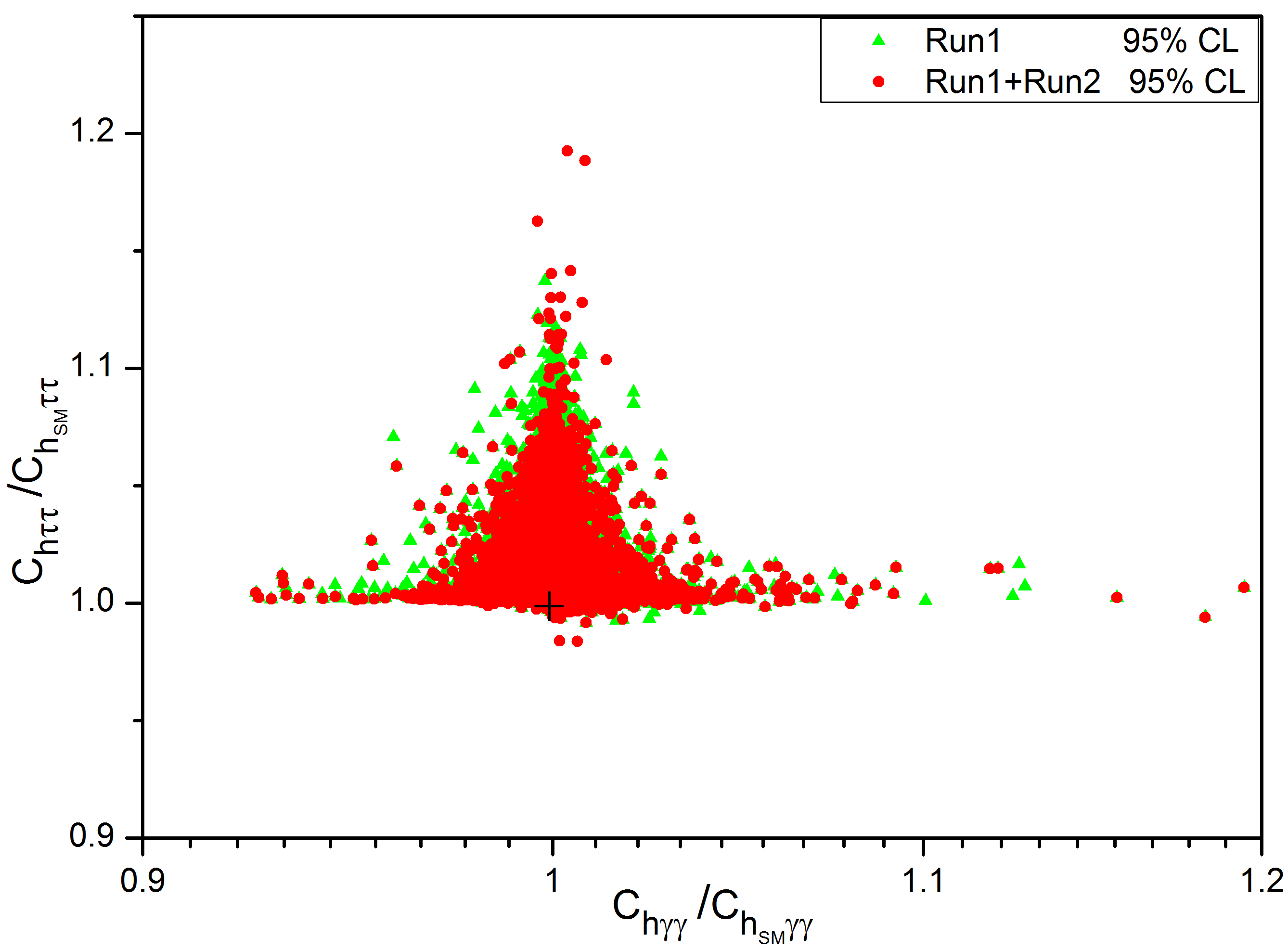}
  \includegraphics[width=5.4cm]{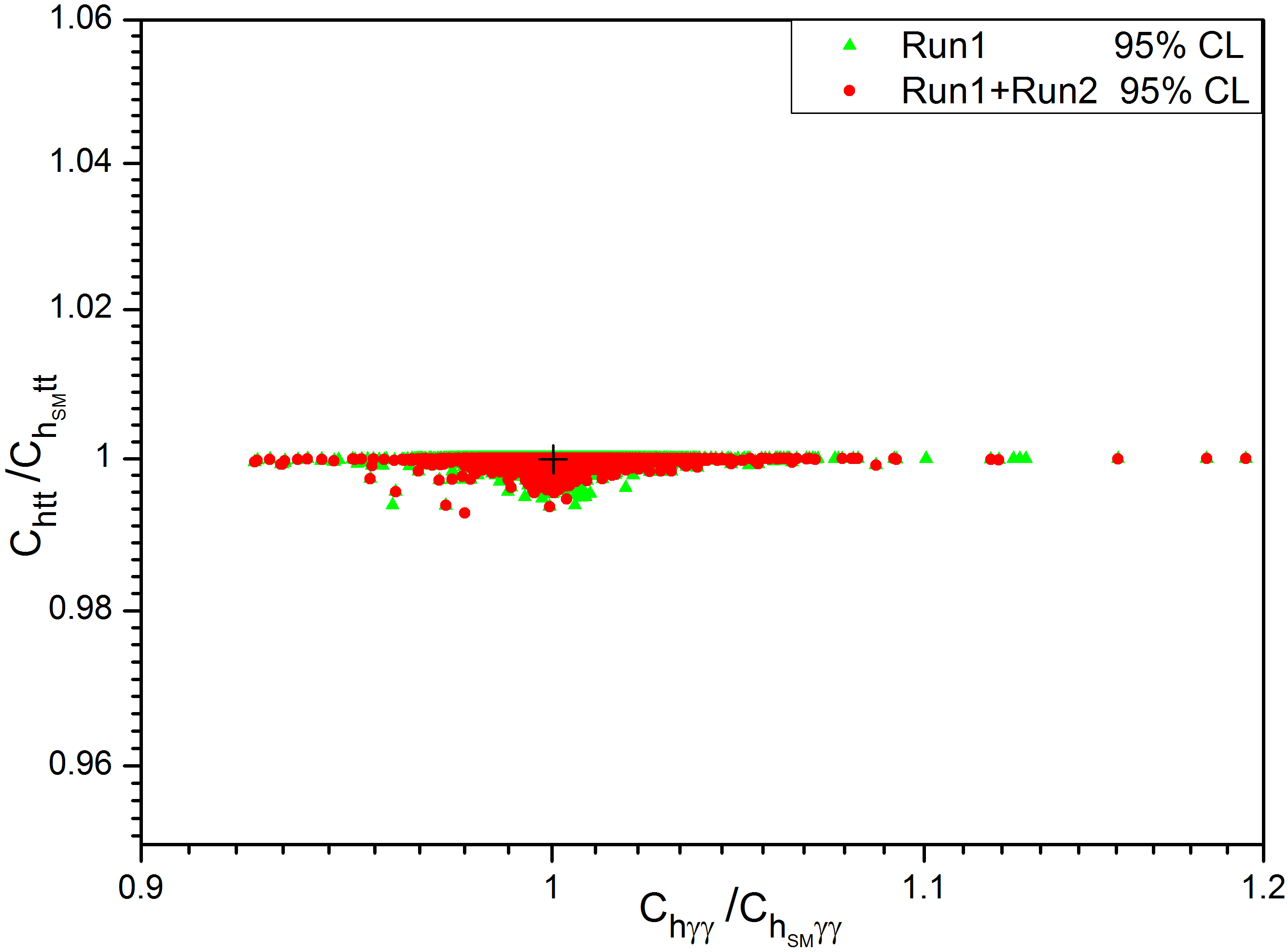}
  \includegraphics[width=5.4cm]{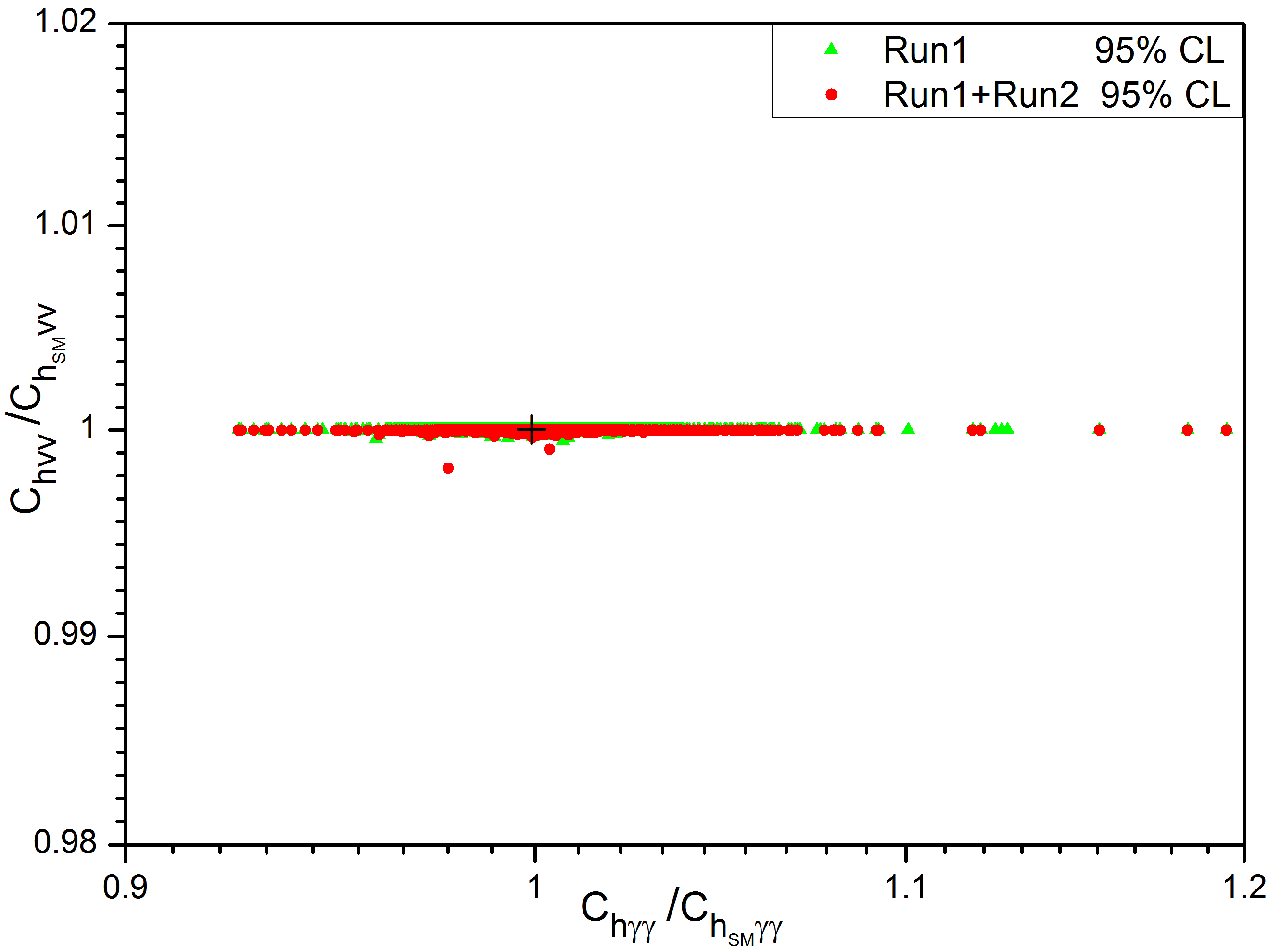}
\vspace{-1.0cm}
\caption{Same as Fig.\ref{higgs-fig2}, but showing the correlations between the couplings 
 normalized to the SM values. }
\label{coupling2}
\end{figure}

\begin{figure}[ht!]
  \centering
  \includegraphics[width=8cm]{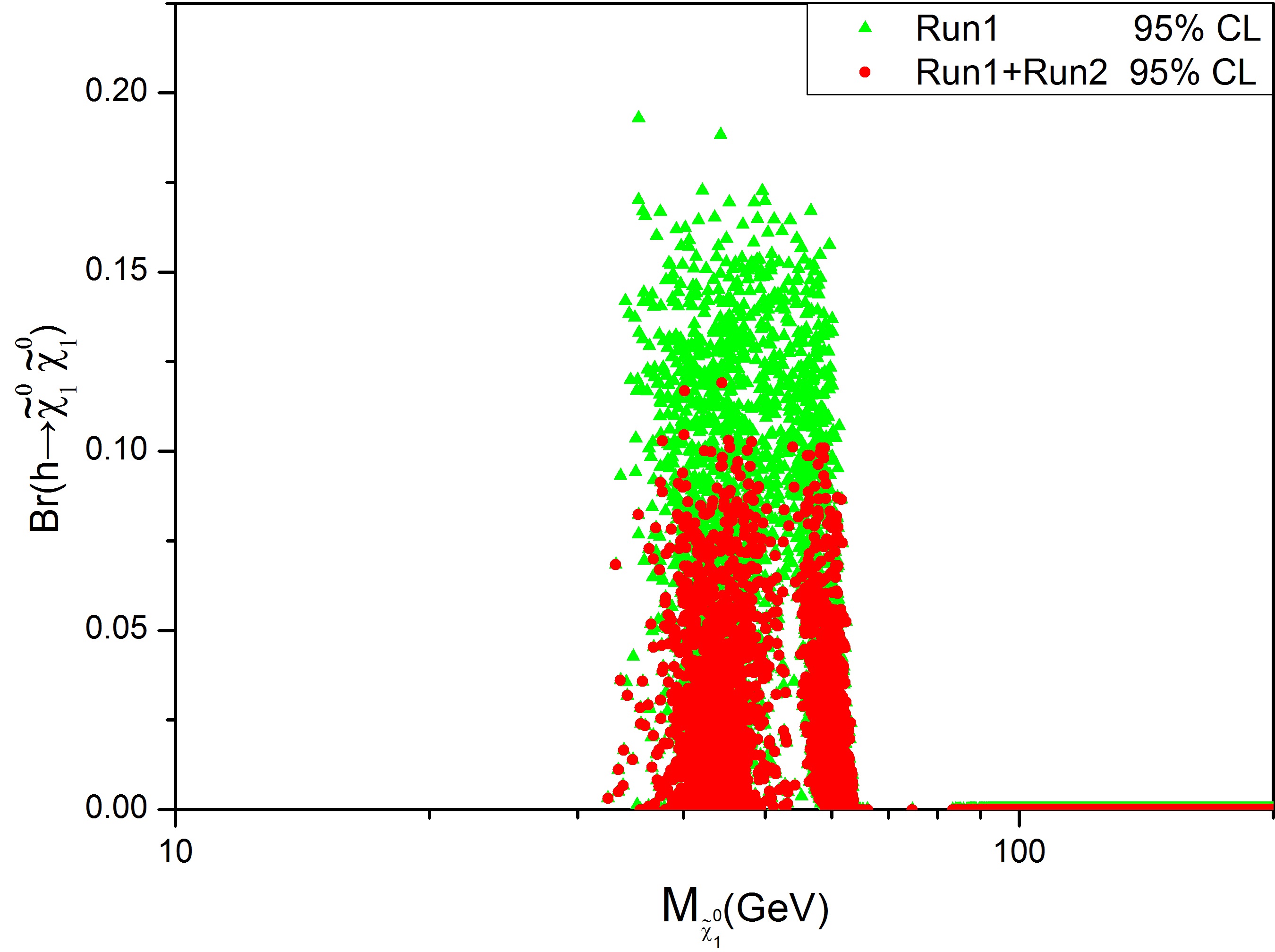}
\vspace{-0.6cm}
\caption{The branching ratio of the decay $h \to \tilde\chi_1^0 \tilde\chi_1^0$ at 95\% CL. }
\label{higgs-fig4}
\end{figure}

In Fig.\ref{coupling2} we show the correlations between the couplings.
As expected, some couplings like $hb\bar{b}$ and $h\tau^+\tau^-$ have strong correlations. 
Since at tree level the $hb\bar{b}$ and  $h\tau^+\tau^-$ couplings are proportional to $\tan\beta$, their deviations with the SM values are sensitive to $\tan\beta$, increasing in magnitude with  the value of $\tan\beta$. For the hvv couplings, their deviations with the SM values are rather small in magnitude and thus not sensitive to $\tan\beta$. Finally, we show the branching ratio of the decay $h \to \tilde\chi_1^0 \tilde\chi_1^0$ in
Fig.\ref{higgs-fig4}.
We see that in the MSSM the branching ratio of this invisible decay is under 10\% currently. 

%%%%%Higgs fig 2  %%%%%%%%%%%%%%%

%%%%%higgs fig.3 %%%%%%%%%%%%%%%%%%%%%

%%%%%higgs fig.4 %%%%%%%%%%%%%%%%%%%%%

\section{Conclusion\label{section4}}
In this work we checked the parameter space of the SM-like Higgs boson 
under current experimental constraints.
We performed a scan over the parameter space
and found that the LHC Run-2 data  can further exclude
a part of parameter space.
In the allowed parameter space,
the gauge couplings of the SM-like Higgs boson $h$ rather approach to the SM values,
while its Yukawa couplings $hb\bar{b}$ and $h\tau^+\tau^-$ can still differ
from the SM predictions by several tens percent.
The deviation of $ht\bar{t}$ coupling from the SM value is below 2\%.
The loop-induced $hgg$ and $h\gamma\gamma$ couplings can differ from the SM values
significantly by several tens percent.
The neutralino dark matter is mostly bino-like and the branching ratio
of the SM-like Higgs decay to dark matter is roughly under 10\%.

Give the present status  of the SM-like Higgs boson in the MSSM, the deviations of its
gauge coupling $hZZ$ from the SM value is almost impossible to detect at CEPC or FCC-ee.
Since its Yukawa couplings $hb\bar{b}$ and $h\tau^+\tau^-$ as well as loop-induced
$h\gamma\gamma$ and  $hgg$ couplings can deviate from their SM values by several tens percent
(at the end of Run-2 or HL-LHC the deviations of these couplings will be further constrained),
it is hopeful to detect such deviations or exclude a crucial part of
the MSSM parameter space in case of unobservation.
Of course, the measurement of the Higgs  invisible decay can also cover some part of
the MSSM parameter space.
The  $ht\bar{t}$ coupling cannot be directly measured at CEPC due to the limited collision
energy, whose precise measurement can be performed at the ILC \cite{ILC}.

\section*{Acknowledgement}
We thank Tim Stefaniak, Guang Hua Duan, Jie Ren and Yang Zhang for technical supports 
and helpful discussions. 
This work was supported by the National Natural Science Foundation of China (NNSFC) 
under grant No. 11375001, by the CAS Key Research Program of Frontier Sciences and 
by a Key R\&D Program of Ministry of Science and Technology of China 
under number 2017YFA0402200-04.

\end{document}